\documentclass{bmcart}

\usepackage{amsthm,amsmath}
\usepackage{braket}
\usepackage[utf8]{inputenc} %unicode support
\usepackage{hyperref}
\usepackage{indentfirst}
\usepackage{multicol}
\usepackage{array}
\usepackage{graphicx}
\usepackage{cancel}
\usepackage{braket}
\usepackage{caption}
\usepackage{subcaption}
\usepackage{float}
\usepackage{amssymb}
\usepackage{bbold}
\usepackage{slashed}
\usepackage{color}
\usepackage{cite}
\usepackage{xfrac}
\usepackage{physics}
\usepackage{siunitx}
\startlocaldefs

\endlocaldefs

\begin{document}

%%% Start of article front matter
\begin{frontmatter}

\begin{fmbox}
\dochead{Research}


\title{Digital quantum simulation of cosmological particle creation with IBM quantum computers}

%%%%%%%%%%%%%%%%%%%%%%%%%%%%%%%%%%%%%%%%%%%%%%
%%                                          %%
%% Enter the authors here                   %%
%%                                          %%
%% Specify information, if available,       %%
%% in the form:                             %%
%%   <key>={<id1>,<id2>}                    %%
%%   <key>=                                 %%
%% Comment or delete the keys which are     %%
%% not used. Repeat \author command as much %%
%% as required.                             %%
%%                                          %%
%%%%%%%%%%%%%%%%%%%%%%%%%%%%%%%%%%%%%%%%%%%%%%
\author[
  addressref={aff1},
  email={marco.diaz@estudiante.uam.es}
]{\inits{M.D.}\fnm{Marco D.} \snm{Maceda}}
\author[
  addressref={aff1},
  email={carlos.sabin@uam.es}   % email address
]{\inits{C.}\fnm{Carlos} \snm{Sabín}}
%
%%%%%%%%%%%%%%%%%%%%%%%%%%%%%%%%%%%%%%%%%%%%%%
%%                                          %%
%% Enter the authors' addresses here        %%
%%                                          %%
%% Repeat \address commands as much as      %%
%% required.                                %%
%%                                          %%
%%%%%%%%%%%%%%%%%%%%%%%%%%%%%%%%%%%%%%%%%%%%%%
 \address[id=aff1]{%                           % unique id
  \orgdiv{Departamento de Física Teórica},             % department, if any
  \orgname{Universidad Autónoma de Madrid},          % university, etc
 \postcode{28049},
  \city{Madrid},              
  \cny{Spain}                 
}

 \address[id=aff2]{%                           % unique id
  \orgdiv{Departamento de Física Teórica},             % department, if any
  \orgname{Universidad Autónoma de Madrid},          % university, etc
 \postcode{28049},
  \city{Madrid},              
  \cny{Spain}                 
}

%%%%%%%%%%%%%%%%%%%%%%%%%%%%%%%%%%%%%%%%%%%%%%
%%                                          %%
%% Enter short notes here                   %%
%%                                          %%
%% Short notes will be after addresses      %%
%% on first page.                           %%
%%                                          %%
%%%%%%%%%%%%%%%%%%%%%%%%%%%%%%%%%%%%%%%%%%%%%%

%%\note{Sample of title note}     % note to the article
%\note[id=n1]{Equal contributor} % note, connected to author

\end{fmbox}% comment this for two column layout

%%%%%%%%%%%%%%%%%%%%%%%%%%%%%%%%%%%%%%%%%%%%%%%
%%                                           %%
%% The Abstract begins here                  %%
%%                                           %%
%% Please refer to the Instructions for      %%
%% authors on https://www.biomedcentral.com/ %%
%% and include the section headings          %%
%% accordingly for your article type.        %%
%%                                           %%
%%%%%%%%%%%%%%%%%%%%%%%%%%%%%%%%%%%%%%%%%%%%%%%

\begin{abstractbox}

\begin{abstract} % abstract

 We use digital quantum computing to simulate the creation of particles in a dynamic spacetime. We consider a system consisting of a minimally coupled massive quantum scalar field in a spacetime undergoing homogeneous and isotropic expansion, transitioning from one stationary state to another through a brief inflationary period. We simulate two vibration modes, positive and negative for a given field momentum, by devising a quantum circuit that implements the time evolution. With this circuit, we study the number of particles created after the universe expands at a given rate, both by simulating the circuit and by actual experimental implementation on IBM quantum computers, consisting of hundreds of quantum gates. We find that state-of-the-art error mitigation techniques are useful to improve the estimation of the number of particles and the fidelity of the state.

%\parttitle{First part title} %if any
%Text for this section.

%\parttitle{Second part title} %if any
%Text for this section.
\end{abstract}

%%%%%%%%%%%%%%%%%%%%%%%%%%%%%%%%%%%%%%%%%%%%%%
%%                                          %%
%% The keywords begin here                  %%
%%                                          %%
%% Put each keyword in separate \kwd{}.     %%
%%                                          %%
%%%%%%%%%%%%%%%%%%%%%%%%%%%%%%%%%%%%%%%%%%%%%%


% MSC classifications codes, if any
%\begin{keyword}[class=AMS]
%\kwd[Primary ]{}
%\kwd{}
%\kwd[; secondary ]{}
%\end{keyword}

\end{abstractbox}
%
%\end{fmbox}% uncomment this for two column layout

\end{frontmatter}

\section{Introduction}

 Gravity is probably a quantum field \cite{why_quantum_gravity_Woodard_2009,Bose_2017} but we don't know yet how to completely fit it into Quantum Field Theory. Standard quantization of General Relativity leads to a nonrenormalizable theory and alternate approaches are far from making falsifiable predictions.

 We can take a more modest path, where spacetime is kept as a classical background and is coupled to the quantum fields. This semiclassical approximation is known as Quantum Field Theory in Curved Spacetime (QFTCS). In this way, we maintain a classical but curved gravitational background, which serves as the stage on which we place the matter fields, which are quantized.
The most famous application of QFTCS is probably Black Hole Thermodynamics, which includes Hawking Radiation \cite{hawking_radiation_Hawking1975}. Other examples are the Unruh Effect \cite{unruh_effect_PhysRevD.14.870}, Hawking's Chronology Protection Conjecture \cite{chronology_protection_PhysRevD.46.603} or cosmological particle creation in expanding spacetimes\cite{parkerPhysRev.183.1057,parkerIIPhysRevD.3.346,parker1976thermal}.

However, these predictions are extremely hard to check experimentally, which suggests the interest of using simulators. For instance, the simulation of Hawking radiation has been successfully achieved in the laboratory with Bose-Einstein condensates \cite{steinhauer} and many theoretical proposals with superconducting circuits or other quantum setups can be found in the literature \cite{quantum_vacuum_hawking_radiation_RevModPhys.84.1,hu2019quantum}. In particular, cosmological models of expanding spacetimes has been also simulated in the laboratory \cite{steinhauer2022analogue,viermann2022quantum}. In comparison to these analogue quantum simulations, it seems that the possibility of digital quantum simulation of QFTCS has been much less explored.

Digital quantum computers have undergone spectacular improvements over the last years, quickly progressing from few-qubit academic demonstrations to the current setups with tens or hundreds of qubits. Despite impressive improvements in the error rates, the current scenario can still be characterized as the Noisy Intermediate-Scale Quantum (NISQ) era \cite{preskill2018quantum}, where on the one hand quantum error correcting codes are still not operational and thus the lack of fault tolerance prevents the elusive dream of universal quantum computing, while on the other hand useful particular tasks beyond the classical computer capabilities are still out of reach, other than tailor-made examples with no clear application -the quantum supremacy paradigm \cite{arute2019quantum,wu2021strong,brooks2019beyond}. However, it has been recently been claimed that the use of error mitigation \cite{ying,temme,kandala} techniques -instead of quantum error-correcting codes- can make current NISQ computers give reliable and useful estimations of expectation values of observables of interest beyond the capabilities of classical computer- a notion called quantum utility \cite{kim2023evidence}. In this work, we show that a current quantum computer can provide an estimation of the number of particles generated in a model of expanding spacetime. We present results for the number of particles generated and the fidelity of the state in a quantum circuit containing hundreds of quantum gates, which simulates the time evolution of two modes of a quantum field under the spacetime dynamics. We use both IBM simulators and real quantum computers. Despite the large amount of gates, we show that the use of zero-noise extrapolation \cite{kim2023evidence} allows us to reach interesting predictions even in the actual quantum devices with their current levels of noise.

In section \ref{secmodel} we explain the details of the model that we digitize in section \ref{secdig}. We present the results of simulations and experiments in sections \ref{secnum} and \ref{secfid}.

\section{Model}
\label{secmodel}

We will follow the toy model provided in \cite{birreldavis}. The infinitesimal line element $ds$, whose coefficients determine the metric of the spacetime, in a Friedman-Lemaitre-Robertson-Walker universe is given by\cite{friedmann1, friedmann2, lemaitre, robertson1, robertson2, robertson3, walker}:
\begin{equation}
ds^2=dt^2-a^2(t)dx^2
\end{equation}
where $a(t)$ is the scale factor, which only depends on time.

We can make a change of variables to conformal time, such that $d\eta=\frac{dt}{a(t)}$, so the metric in these coordinates becomes:
\begin{equation}
ds^2=a^2(\eta)\left(d\eta^2-dx^2\right)=C(\eta)\left(d\eta^2-dx^2\right)
\end{equation}
where $C(\eta)=a^2(\eta)$, which is obtained by substituting time $t$ for conformal time $\eta$ through solving the following integral:
\begin{equation}
\eta=\int\frac{dt}{a(t)}
\end{equation}

\begin{figure}
 
\includegraphics[width=0.95\textwidth]{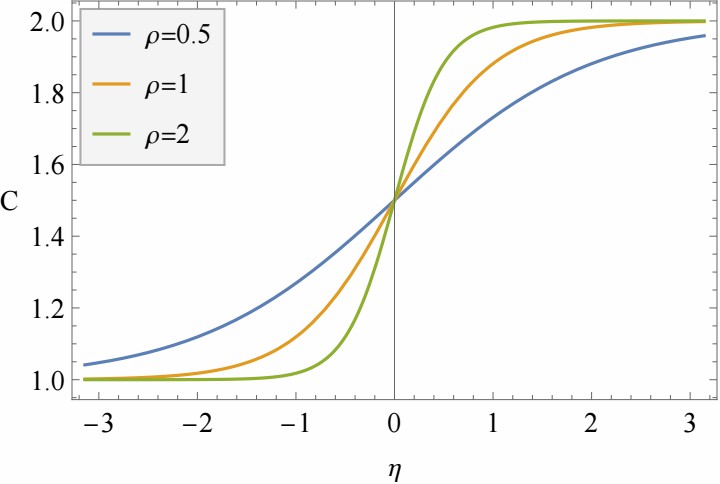}
 
 \caption{Scale factor as a function of time for a universe that expands with a speed \( \rho=0.5 \) (blue), \( \rho=1 \) (orange), and \( \rho=2 \) (green), with \( A=1.5 \) and \( B=0.5 \).}
 \label{fig1}
\end{figure}

We work in a universe that evolves from a stationary state where $C(-\infty)=A-B$ to a new stationary state where $C(\infty)=A+B$ through $C(\eta)=A+B\tanh(\rho\eta)$, where $\rho>0$ is a constant that codifies the transition velocity (see Figure \ref{fig1}). The universe expands if $B>0$, contracting otherwise. The value of $C(\eta)$ must always be positive, so $A>|B|$.

On this gravitational background, we place a free scalar field with mass $m$, that is, it evolves over time through the Klein-Gordon equation\cite{klein, gordon}:

% \begin{figure}
% \includegraphics[width=0.95\textwidth]{fig1.jpg}
% \caption{Scale factor as a function of time for a universe that expands with a speed \( \rho=0.5 \) (blue), \( \rho=1 \) (orange), and \( \rho=2 \) (green), with \( A=1.5 \) and \( B=0.5 \).} 
% \label{fig1}
% \end{figure}
% \captionsetup[figure]{font=footnotesize,labelfont=footnotesize}
\begin{equation}
\frac{\partial^2\phi}{\partial\eta^2}-\frac{\partial^2\phi}{\partial x^2}+C(\eta)m^2\phi=0
\end{equation}

This partial differential equation can be solved by separation of variables, with a solution for each mode of the form:
\begin{equation}
\phi_k(\eta, x)=\frac{1}{\sqrt{2\pi}}e^{ikx}\chi_k(\eta)
\end{equation}
with $\chi_k(\eta)$ satisfying:
\begin{equation}
\frac{d^2\chi_k}{d\eta^2}+\left(k^2+C(\eta)m^2\right)\chi_k=0
\end{equation}
whose solution is given by hypergeometric functions.

We are particularly interested in the behavior of the field in the limits $\eta\to\pm\infty$, which is given by:
\begin{equation}
\phi_k(\eta\to-\infty, x)\to\frac{1}{\sqrt{4\pi\omega_{in}}}e^{i\left(kx-\omega_{in}\eta\right)}
\end{equation}
\begin{equation}
\phi_k(\eta\to\infty, x)\to\frac{1}{\sqrt{4\pi\omega_{out}}}e^{i\left(kx-\omega_{out}\eta\right)}
\end{equation}
where $\omega_{in}=\sqrt{k^2+m^2(A-B)}$ and $\omega_{out}=\sqrt{k^2+m^2(A+B)}$. These two limits are of special interest because $\omega_{out}$, the frequency of the free field after the expansion, is the frequency that will appear in the Hamiltonian governing the time evolution of the field after the universe expands. Additionally, together with $\omega_{in}$, these are the frequencies that are needed to relate the vibrational modes of the field before and after the expansion, which is crucial for computing the number of particles created by the expansion of the universe.

Clearly, the vibration modes of the field are not the same before and after a short period of inflation. Initially, the field is stationary in the vacuum state $\ket{0, in}$. After the inflationary evolution, however, this state is not going to be an eigenstate of the time evolution anymore, so we could detect a number of particles different from zero and study their evolution over time.

We have chosen this model because the vibration modes of the field before inflation can be easily related to those after inflation, in the sense that the vibrations with momentum $k$ mix only with their opposite $-k$, but not with other configurations with different $|k|$. The Bogoliubov transformations that relate the vibration modes $\phi^{in}$ and $\phi^{out}$ are given by:
\begin{equation}
\phi_k^{in} = \alpha_k \phi_k^{out} + \beta_k \phi_{-k}^{out*}
\end{equation}
where the coefficients $\alpha$ and $\beta$ are complex functions of $\omega_{in}$, $\omega_{out}$ and $\rho$ \cite{birreldavis}:
\begin{equation}
\alpha_k = \sqrt{\frac{\omega_{out}}{\omega_{in}}} \frac{\Gamma\left(1-i\frac{\omega_{in}}{\rho}\right)\Gamma\left(-i\frac{\omega_{out}}{\rho}\right)}{\Gamma\left(-i\frac{\omega_+}{\rho}\right)\Gamma\left(1-i\frac{\omega_+}{\rho}\right)}, \quad\quad\quad\quad \beta_k = \sqrt{\frac{\omega_{out}}{\omega_{in}}} \frac{\Gamma\left(1-i\frac{\omega_{in}}{\rho}\right)\Gamma\left(i\frac{\omega_{out}}{\rho}\right)}{\Gamma\left(i\frac{\omega_-}{\rho}\right)\Gamma\left(1+i\frac{\omega_-}{\rho}\right)} \label{eq:alpha
}
\end{equation}
with $\omega_\pm=\frac{1}{2}\left(\omega_{out}\pm\omega_{in}\right)$.

The reason why we are using this toy model to simulate the cosmological particle production is because its simplicity allows us to write an analytical expression for the Bogoliubov coefficients, which relate the vibrational modes of the field, and its respective quantized observables, before and after the expansion. For other models whose relationship between modes is not known, it is not possible to perform such simulations. Furthermore, in more general models of QFTCS, the number of particles is not well-defined, depending on the observer measuring it. A classical example of this is black hole thermal radiation, which is zero for a near-horizon freely falling observer, but non-zero for a stationary one. In contrast, our model allows us choose an observer who initially detects the vacuum and study the particle creation from its perspective.

When quantizing the field, we use the annihilation and creation operators $a_+$ and $a_+^\dagger$ for the $in$ modes of positive $k$, and $a_-$ and $a_-^\dagger$ for the $in$ modes of negative $k$. These creation and annihilation operators are related to the annihilation and creation operators of the $out$ modes, which we call $b_+$, $b_+^\dagger$, $b_-$ and $b_-^\dagger$, using the same Bogoliubov coefficients\cite{bogoliubov, valatin}:
\begin{equation}
b_+ = \alpha a_+ + \beta^* a_-^\dagger \quad\quad\quad\quad b_+^\dagger = \alpha^* a_+^\dagger + \beta a_-
\end{equation}
\begin{equation}
b_- = \alpha a_- + \beta^* a_+^\dagger \quad\quad\quad\quad b_-^\dagger = \alpha^* a_-^\dagger + \beta a_+
\end{equation}

Now, the generator of the time evolution of these two positive and negative modes is the Hamiltonian:
\begin{equation}
\begin{split}
\hat{\mathcal{H}} &= \hbar \omega_{out} \left( b_+^\dagger b_+ + b_-^\dagger b_- \right) \\
&= \hbar \omega_{out} \left( \alpha \alpha^* \left( a_+^\dagger a_+ + a_-^\dagger a_- \right) + \beta \beta^* \left( a_+ a_+^\dagger + a_- a_-^\dagger \right) + 2 \alpha^* \beta a_+^\dagger a_-^\dagger + 2 \alpha \beta^* a_+ a_- \right)
\end{split}
\end{equation}

In the interaction picture, an initial vacuum state will evolve under the interaction hamiltonian:
\begin{equation}
\mathcal{H}_{int} = 2 \hbar \omega_{out} \alpha \beta^* a_+ a_- + h.c. \label{eq:inthamiltonian
}
\end{equation}

Thus the time evolution operator in this picture is:
\begin{equation}
U = e^{i \omega_{out} t \left( 2 \alpha \beta^* a_+ a_- + h.c. \right)} \label{evol}
\end{equation}

 We are interested in the number of particles -as defined before the expansion by the in operators- created after the expansion as a function of the expansion rate $\rho$.

This type of particle creation process due to a non-flat metric can be interpreted as if the field were in thermodynamic equilibrium with a thermal source\cite{hawking, unruh}. Therefore, the state predicted by quantum field theory for the time evolution operator \ref{evol} applied to an initial vacuum is a state well-known in quantum optics -two-mode squeezed state \cite{PhysRevA.31.3093} where $2 \alpha \beta^* \omega_{out} t$ is the squeezing parameter. When the trace over one mode is taken over a two-mode squeezed state, a single-mode thermal state with squeezing-dependent temperature is obtained \cite{PhysRevA.36.3464}, giving rise to:
\begin{equation}
\rho = \frac{1}{\cosh^2 \left( 2 \left| \alpha \right| \left| \beta \right| \omega_{out} t \right)} \sum_{n=0}^\infty \tanh^{2n} \left( 2 \left| \alpha \right| \left| \beta \right| \omega_{out} t \right) \ket{n} \bra{n} \label{termicexpansion}
\end{equation}

Therefore, the expected number of particles created can be calculated as:
\begin{equation}
\expval{n} = \sinh^2\left(2\left|\alpha\right|\left| \beta\right|\omega_{out}t\right) \label{ninf}
\end{equation}

As we will see below, we will restrict ourselves to only one created particle per mode. The state restricted to one excitation can be obtained by taking the first two terms of the sum in \ref{termicexpansion} and renormalizing, giving rise to:
\begin{equation}
\expval{n} = \frac{\tanh^2 \left( 2 \left| \alpha \right| \left| \beta \right| \omega_{out} t \right)}{1 + \tanh^2 \left( 2 \left| \alpha \right| \left| \beta \right| \omega_{out} t \right)} \label{nuno}
\end{equation}

\section{Digitization}
\label{secdig}

To simulate the system, we assign the first two qubits to the activation of the positive mode, and the last two to the negative mode \cite{somma}. Therefore, we assign $\ket{0101}$ to the state $\ket{0, in}$, the qubit $\ket{1001}$ to the state $\ket{1_+, in}$, the qubit $\ket{0110}$ to the state $\ket{1_-, in}$, and the qubit $\ket{1010}$ to the state $\ket{1_+1_-, in}$. This means that we are restricting ourselves to only one excitation per mode.

In this qubit basis, the creation and annihilation operators map to:
\begin{equation}
a_+^\dagger=\sigma_+^0\sigma_-^1\,\,\,\,\,\,\,\,\,\,\,\,\,\,\,
a_+=\sigma_-^0\sigma_+^1
\end{equation}
\begin{equation}
a_-^\dagger=\sigma_+^2\sigma_-^3\,\,\,\,\,\,\,\,\,\,\,\,\,\,\,a_-=\sigma_-^2\sigma_+^3
\end{equation}
where the matrices $\sigma_\pm$ are given by:
\begin{equation}
\sigma_\pm=\frac{1}{2}\left(\sigma_x\pm i\sigma_y\right)\label{pm}
\end{equation}

Under this qubit mapping, the interaction Hamiltonian (\ref{eq:inthamiltonian
}) is written as:
\begin{equation}
\mathcal{H}_{int}=2\hbar\omega_{out}\left(\alpha\beta^*\sigma_-^0\sigma_+^1\sigma_-^2\sigma_+^3+\alpha^*\beta\sigma_+^0\sigma_-^1\sigma_+^2\sigma_-^3\right)
\end{equation}

Using (\ref{pm}), we write the operators in terms of Pauli matrices:
\begin{equation}
\sigma_-^0\sigma_+^1\sigma_-^2\sigma_+^3=\frac{1}{16}(A+iB);\quad\quad \sigma_+^0\sigma_-^1\sigma_+^2\sigma_-^3=\frac{1}{16}(A-iB),
\end{equation}
where
\begin{eqnarray}
A&=& \sigma_x^0\sigma_x^1\sigma_x^2\sigma_x^3+\sigma_y^0\sigma_y^1\sigma_y^2\sigma_y^3+\sigma_x^0\sigma_x^1\sigma_y^2\sigma_y^3-\sigma_x^0\sigma_y^1\sigma_x^2\sigma_y^3+\nonumber\\ & &\sigma_y^0\sigma_x^1\sigma_x^2\sigma_y^3+\sigma_x^0\sigma_y^1\sigma_y^2\sigma_x^3-
\sigma_y^0\sigma_x^1\sigma_y^2\sigma_x^3+\sigma_y^0\sigma_y^1\sigma_x^2\sigma_x^3,
\end{eqnarray}
and
\begin{eqnarray}
B&=& \sigma_x^0\sigma_x^1\sigma_x^2\sigma_y^3-\sigma_x^0\sigma_x^1\sigma_y^2\sigma_x^3+\sigma_x^0\sigma_y^1\sigma_x^2\sigma_x^3-\sigma_y^0\sigma_x^1\sigma_x^2\sigma_x^3+\nonumber\\ & &\sigma_x^0\sigma_y^1\sigma_y^2\sigma_y^3-\sigma_y^0\sigma_x^1\sigma_y^2\sigma_y^3+\sigma_y^0\sigma_y^1\sigma_x^2\sigma_y^3-\sigma_y^0\sigma_y^1\sigma_y^2\sigma_x^3.
\end{eqnarray}

Therefore, the temporal evolution operator would be given by the complex exponential of the following interaction Hamiltonian:
\begin{equation}
    \mathcal{H}_{int}=\frac{\hbar\omega_{out}}{4}\left(\Re(\alpha\beta^*)A-\Im(\alpha\beta^*)B\right).
  \label{temp}
\end{equation}

Note that all terms within $A$ and $B$ commute with each other, while the commutator of A and B  only has two non-zero diagonal terms that merely add an irrelevant phase to the initial vacuum state. 
Therefore the decomposition into exponential products is exact up to second order in $\alpha$ and $\beta$ and no further use of Trotter techniques is required.
%At third order, the decomposition has an error of $\frac{16}{3}\omega^3t^3\Im\left(\alpha\beta^*\right)\Re\left(\alpha\beta^*\right)\left(\Re\left(\alpha\beta^*\right)+2i\Im\left(\alpha\beta^*\right)\right)$ which, for the simulation parameters we have chosen, reaches a maximum value of $0.000252+0.0000134i$, so we remain in a reasonable region within the perturbative approximation.

Thus by factorizing the time evolution we get something of the form:
\begin{equation}
U=e^{-\frac{i}{4}\omega_{out}t\mathrm{Re}(\alpha\beta^*)\sigma_x^0\sigma_x^1\sigma_x^2\sigma_x^3}e^{-\frac{i}{4}\omega_{out}t\mathrm{Re}(\alpha\beta^*)\sigma_y^0\sigma_y^1\sigma_y^2\sigma_y^3}\cdots e^{\frac{i}{4}\omega_{out}t\mathrm{Im}(\alpha\beta^*)\sigma_x^0\sigma_x^1\sigma_x^2\sigma_y^3}\cdots
\end{equation}

Now we have to decompose these multiqubit interactions into single and two-qubit gates. To this aim we resort to the following decomposition\cite{sabin1, sabin2, sabin3, sabin4}:
\begin{equation}
\sigma_x^0\sigma_x^1\sigma_x^2\sigma_x^3=e^{-i\frac{\pi}{4}\sigma_z^0\sigma_x^3}e^{-i\frac{\pi}{4}\sigma_z^0\sigma_x^2}e^{-i\frac{\pi}{4}\sigma_z^0\sigma_x^1}e^{-i\frac{\pi}{4}\sigma_x^0}\left(-\sigma_z^0\right)e^{i\frac{\pi}{4}\sigma_x^0}e^{i\frac{\pi}{4}\sigma_z^0\sigma_x^1}e^{i\frac{\pi}{4}\sigma_z^0\sigma_x^2}e^{i\frac{\pi}{4}\sigma_z^0\sigma_x^3},\label{H0}
\end{equation}

%\begin{figure}
 %\caption{Circuit performing the corresponding temporal evolution for the operator $\sigma_x^0\sigma_x^1\sigma_y^2\sigma_y^3$.}
 %\label{fig2}
%\end{figure}

\noindent and, in case we have a gate $y$ instead of a gate $x$ acting on qubit $i$, we introduce a rotation matrix to change the $y$ into $x$ as follows:
\begin{equation}
\sigma_y^i=e^{-i\frac{\pi}{4}\sigma_z^i}\sigma_x^ie^{i\frac{\pi}{4}\sigma_z^i}\label{xy}
\end{equation}

The two-qubit gates can be translated to CNOT gates, by using:
\begin{equation}
e^{i\frac{\pi}{4}\sigma_z^i\sigma_x^j}=e^{i\frac{\pi}{4}\sigma_z^i}e^{i\frac{\pi}{4}\sigma_x^j}CNOT^{ij}\label{CNOT}
\end{equation}
except for a global phase that we disregard.

Now that we have the operator $\sigma_a^0\sigma_b^1\sigma_c^2\sigma_d^3$ written in the following form:
\begin{equation}
\sigma_a^0\sigma_b^1\sigma_c^2\sigma_d^3=M^\dagger\left(-\sigma_z^0\right)M,
\end{equation}
with $M$ a unitary matrix, which is given by the equations \ref{H0} and \ref{xy}, we can take exponentials in both sides to find:
\begin{equation}
e^{ik\sigma_a^0\sigma_b^1\sigma_c^2\sigma_c^3}=M^\dagger e^{-ik\sigma_z^0}M
\end{equation}
where $k$ is the coefficient accompanying that operator in the temporal evolution, which can be $-\frac{1}{4}\omega_{out}t\mathrm{Re}(\alpha\beta^*)$ or $\frac{1}{4}\omega_{out}t\mathrm{Im}(\alpha\beta^*)$.

The operator left in the middle is simply a qubit rotation around the $z$ axis whose rotation angle depends on the coefficients $\alpha$ and $\beta$ the therefore on the expansion rate $\rho$.

\begin{figure}
\includegraphics[width=0.95\textwidth]{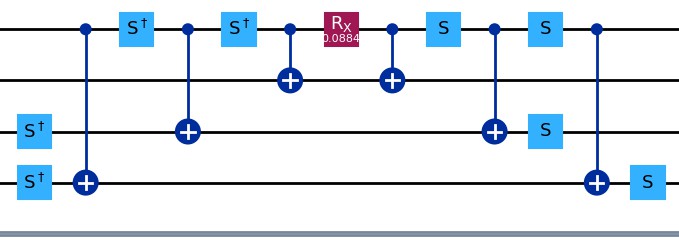}
\label{fig2}
\caption{Circuit performing the corresponding temporal evolution for the operator $\sigma_x^0\sigma_x^1\sigma_y^2\sigma_y^3$.}
\end{figure}
\captionsetup[figure]{font=footnotesize,labelfont=footnotesize}

For example, for the case $\sigma_x^0\sigma_x^1\sigma_y^2\sigma_y^3$, the contribution to the temporal evolution is:
\begin{eqnarray}
U&=&e^{\frac{i}{4}\omega_{out}t\mathrm{Re}(\alpha\beta^*)\sigma_x^0\sigma_x^1\sigma_y^2\sigma_y^3}=e^{-i\frac{\pi}{4}\sigma_z^2}e^{-i\frac{\pi}{4}\sigma_z^3}e^{-i\frac{\pi}{4}\sigma_z^0\sigma_x^3}e^{-i\frac{\pi}{4}\sigma_z^0\sigma_x^2}e^{-i\frac{\pi}{4}\sigma_z^0\sigma_x^1}e^{-i\frac{\pi}{4}\sigma_x^0}\nonumber\\& &e^{\frac{i}{4}\omega_{out}t\mathrm{Re}(\alpha\beta^*)\sigma_z^0}e^{i\frac{\pi}{4}\sigma_x^0}e^{i\frac{\pi}{4}\sigma_z^0\sigma_x^1}e^{i\frac{\pi}{4}\sigma_z^0\sigma_x^2}e^{i\frac{\pi}{4}\sigma_z^0\sigma_x^3}e^{i\frac{\pi}{4}\sigma_z^3}e^{i\frac{\pi}{4}\sigma_z^2}=S_3S_2CNOT_{03}S_0\nonumber\\& &CNOT_{02}S_0CNOT_{01}RX_0\left(-\frac{1}{2}\omega_{out}t\mathrm{Re}(\alpha\beta^*)\right)CNOT_{01}S^\dagger_0CNOT_{02}\nonumber\\& &S^\dagger_0CNOT_{03}S^\dagger_2S^\dagger_3
\end{eqnarray}
where, in the last step, equation \ref{CNOT} has been used, many simplifications have been made, and it has been written in terms of matrices $S$ and $RX$ in order to express it in the language of IBM quantum computers.

\begin{figure}[h!]
 \includegraphics[width=0.95\textwidth]{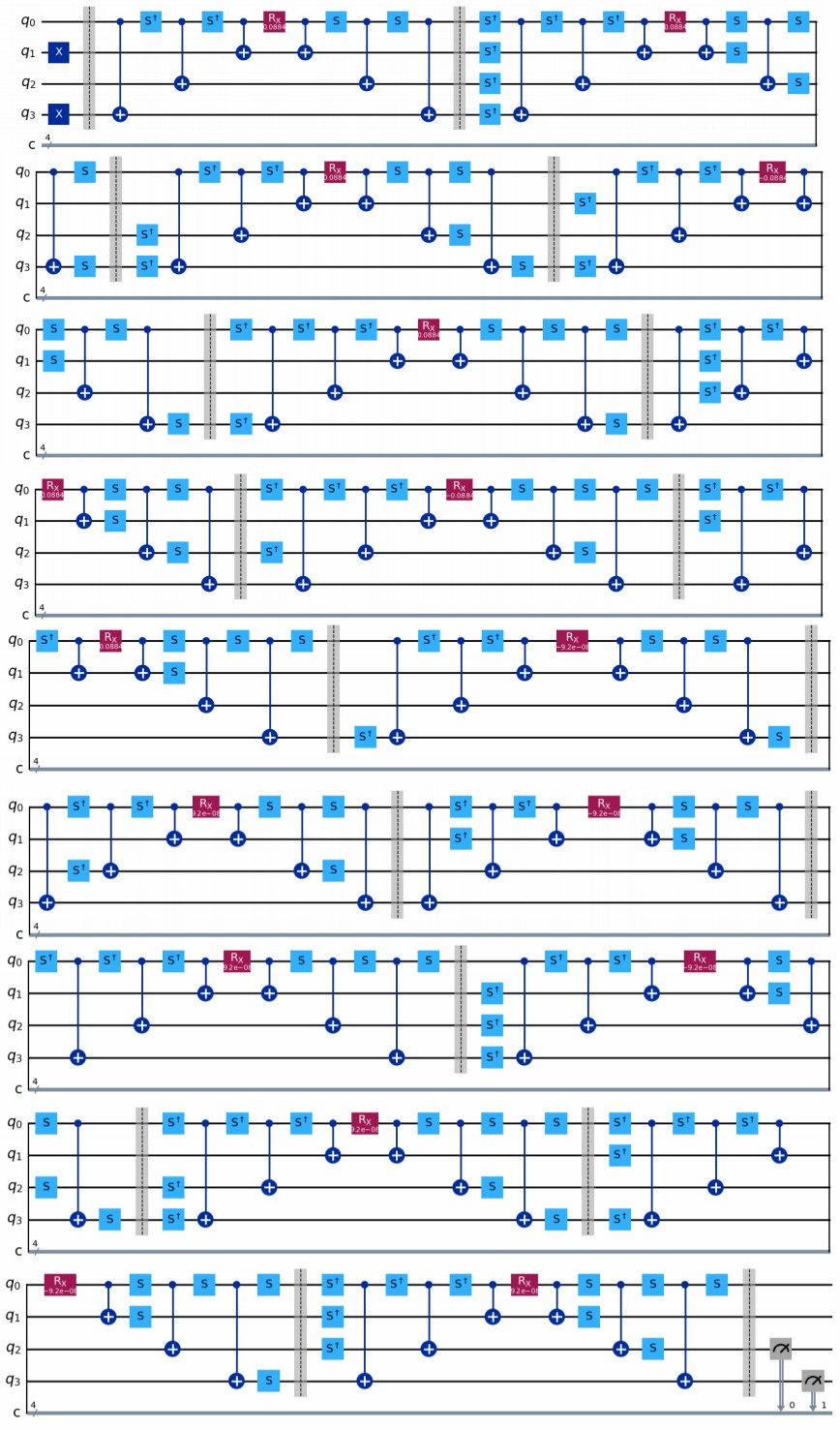}
 \caption{Complete circuit designed to measure the number of thermal particles created by a homogeneous and isotropic expansion of the universe.}
 \label{fig3}
\end{figure}

This circuit can be observed in Figure \ref{fig2}. In addition to this, there are 15 other similar circuits corresponding to the 16 operators in the equation \ref{temp}. Finally, we  need to place two gates $X_1$ and $X_3$ at the beginning to initialize the state $\ket{0, in}=\ket{0101}$  The complete circuit can be seen in Figure \ref{fig3} including two $z$-axis measurement gates at the end to measure the number of particles in one of the two modes. The latter serves to visualize that we are only measuring the number of particles of one mode, however what we actually do in the experiments is to use the IBM primitive estimator for the computation of four-qubit expectation values and tomographycally reconstruct the probability of excitation of a single mode -see more details in the next section.

%\begin{figure}
 % \caption{Complete circuit designed to measure the number of thermal particles created by a homogeneous and isotropic expansion of the universe.}
  %\label{fig3}
%\end{figure}

 \captionsetup[figure]{font=footnotesize,labelfont=footnotesize}

\section{Results: number of particles}
\label{secnum}

To carry out the simulation, we set $A=1.5$ and $B=0.5$, such that we evolve from a universe with $C=1$ to a universe with $C=2$ with a variable value of $\rho$, ranging from $\rho=0.01$ to $\rho=100$, to study the number of particles generated in a time of $t=1$. Also, we set the mass of the field to $m=1$.

Since the vibration modes do not mix, we study a single excited state of the field, with $|k|=1$, so the vibration frequencies are $\omega_{in}=\sqrt{2}$, $\omega_{out}=\sqrt{3}$, $\omega_+=\frac{1}{2}\left(\sqrt{3}+\sqrt{2}\right)$, and $\omega_-=\frac{1}{2}\left(\sqrt{3}-\sqrt{2}\right)$.

Given that both the positive and negative modes are equivalent, we only measure the number of particles in two out of the four qubits we use. A measurement of $\ket{01}$ in these two qubits represents that we measured the field without any particles, while a measurement of $\ket{10}$ means that we measured a particle in that mode. The states $\ket{00}$ and $\ket{11}$ have no physical meaning. Therefore, the number of particles of, for instance, the second mode would be given by the probabilities of $\ket{1010}$ and $\ket{0110}$.
In order to be able to benefit from error mitigation techniques, we need to use estimator, and therefore we need to estimate the expectation values of the following seven observables $\expval{I\otimes I\otimes I\otimes Z},$ $ \expval{I\otimes I\otimes Z\otimes I},$ $ \expval{I\otimes I\otimes Z\otimes Z},$ $ \expval{Z\otimes Z\otimes I\otimes I},$ $ \expval{Z\otimes Z\otimes I\otimes Z},$ $\expval{Z\otimes Z\otimes Z\otimes I}$ and $\expval{Z\otimes Z\otimes Z\otimes Z}$, which allow the tomographic reconstruction of the aforementioned sum of probabilities.

\begin{figure}
\includegraphics[width=0.95\textwidth]{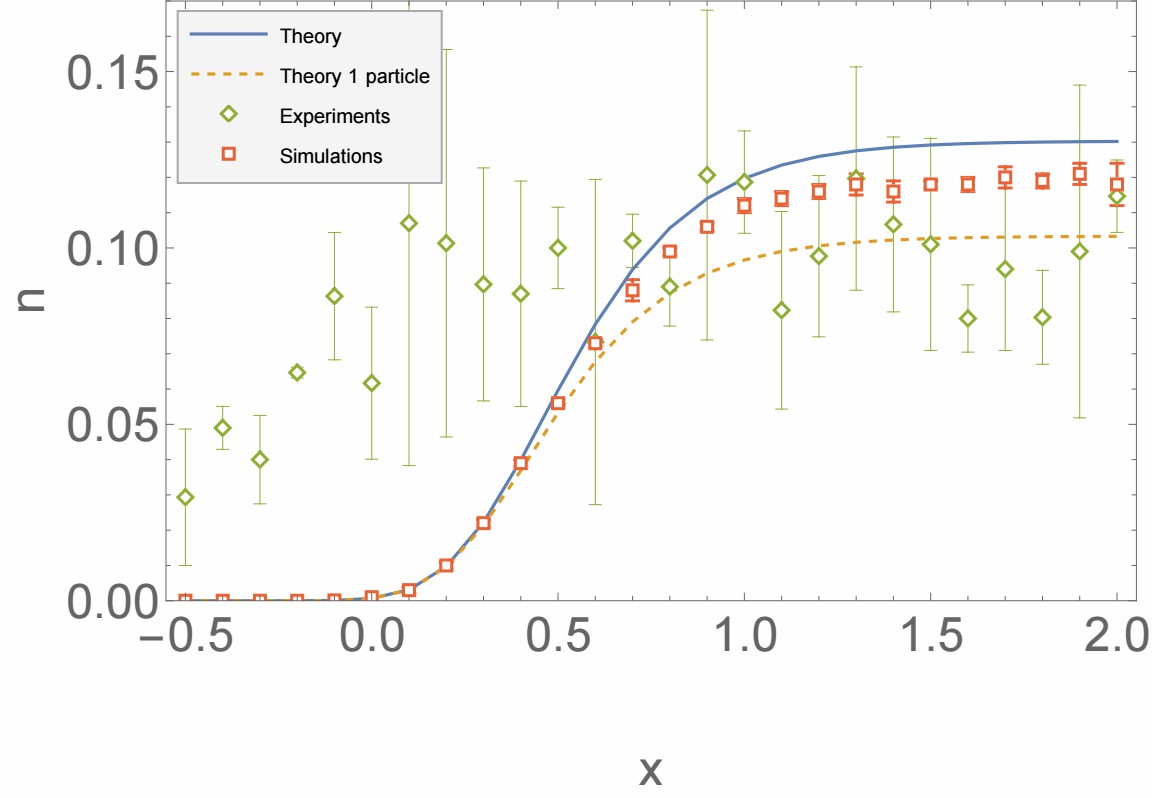}
 \label{fig4}
\caption {Expected value of thermal particles produced by the universe expansion as a function of the expansion rate $\rho=10^x$, both for simulations and experiments, compared to theoretical predictions, with and without constraining to 1 maximum particle per mode.}
 \end{figure}
\captionsetup[figure]{font=footnotesize,labelfont=footnotesize}

To obtain results, we conducted three batches of experiments, each batch consisting in several executions of the circuit over a range of values of $\rho$. First, we performed them with the \textit{AerSimulator} simulator provided by IBM. As can be seen in Figure \ref{fig4}, the simulated values are arranged between the theoretical values for infinite particles and one particle, calculated with equations \ref{ninf} and \ref{nuno}, showing the suitability of our quantum circuit.

Furthermore, we realized actual experiments on the IBM 127-qubit Eagle-processor quantum computer called $ibmq\_osaka$ -which, at the time of the realization of the experiments was one of the computers with the lowest error rates available, see more details below-   with the use of zero-noise extrapolation (ZNE), a state-of-the-art error mitigation technique\cite{ying, temme, kandala}. Results are quite noisy and tend to overestimate the number of particles for negative and small values of $x\equiv\log{\rho}$, but seem to converge to the correct ones for the largest
values of the expansion rate parameter. To obtain more accurate results, it may be necessary to improve the noise removal of the IBM quantum gates or develop better error mitigation techniques.

As a technical remark, we note that we didn't use the highest level of circuit optimization along none of the experiments, since we found that for negative and small values of $x$ the value of the circuit parameters was so low that they were neglected and therefore the experiments were finally just trivial executions of the identity. While this would get us close to the theoretical predictions -which are basically 0 in that parameter range, as can be seen in the Figure- that would of course not be a consequence of any property of the actual quantum hardware. Therefore, we chose to decrease the optimization level, resulting in actual executions of quantum circuits with very small rotation parameters.

\section{Results: fidelity}
\label{secfid}

In order to complement and further illuminate the results for the number of particles, we will now compute the fidelity:
\begin{equation}
F(\rho, \sigma)=\mathrm{Tr}^2\left(\sqrt{\sqrt\rho\sigma\sqrt\rho}\right)\label{fidelity}
\end{equation}
between the state $\rho$, calculated from a second-order Taylor expansion of expression \ref{evol}, and the density matrix $\sigma$ obtained from experimental state tomography. In particular, we focus on the two qubits representing one mode. Fidelity is a measurement of the distance between two quantum states, being 1 if both density matrices represent the same states, and 0 if the states are orthogonal and therefore it can be a useful figure of merit in quantum simulation, by using it to compare the experimental outcome with a theoretical prediction. 
After some algebra, we find that, up to first order to find in $\omega_{out}t$:
\begin{eqnarray}
    F&\approx&\frac{1}{4}\left(1-\expval{I\otimes Z}+\expval{Z\otimes I}-\expval{Z\otimes Z}\right)+\nonumber\\& &\sqrt{(1-\expval{Z\otimes Z})^2-(\expval{I\otimes Z}-\expval{Z\otimes I})^2}|\alpha||\beta|\omega_{out}t.
\end{eqnarray}

\begin{figure}
\includegraphics[width=0.95\textwidth]{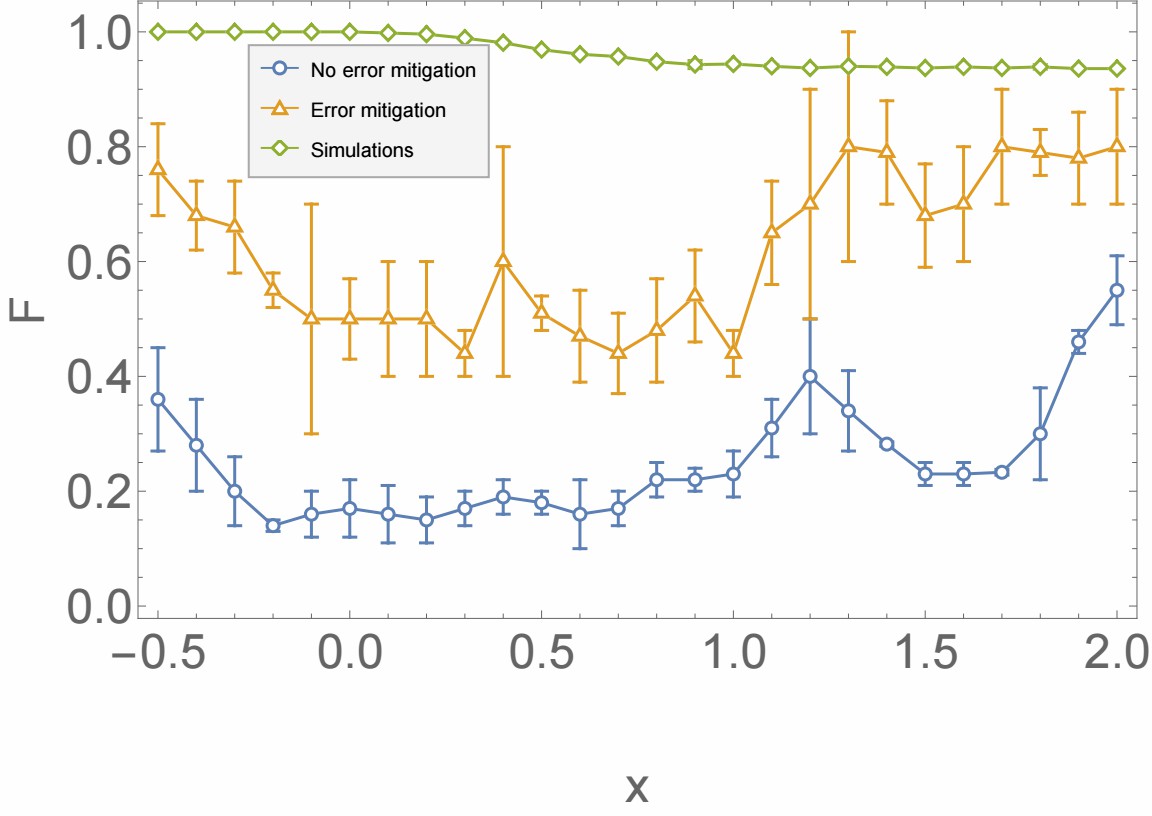}
\label{fig5}
 \caption{Fidelity between the state after executing the circuit and reconstructing it through tomography, both for simulations and actual experiments, with and without the use of zero-noise extrapolation, and the second-order approximation of the theoretical state as a function of the expansion rate $\rho=10^x$.}
 \end{figure}
\captionsetup[figure]{font=footnotesize,labelfont=footnotesize}
Therefore, we only need to measure experimentally the expectation values of three observables: $\expval{I\otimes Z}$, $\expval{Z\otimes I}$, and $\expval{Z\otimes Z}$. We do so for a range of values of $\rho$, with and without error mitigation (ZNE, which characterizes the noise in the experiment and tries to capture its relationship with the results, in order to provide an estimation of the noiseless result), as can be seen in Figure \ref{fig5}. Notice that in order to use estimator, the two-qubit observables would be actually the four-qubit ones, with the identity matrix in the two extra qubits, namely $\expval{I\otimes I\otimes I\otimes Z}$, $\expval{I\otimes I\otimes Z\otimes I}$, and $\expval{I\otimes I\otimes Z\otimes Z}$.

Notice that, without error mitigation, the fidelity is extremely low, as expected from the large number of gates. In particular, to execute our circuit on the IBM Osaka quantum computer, 96 2-qubit gates are used, each with an average associated error of $6.741\cdot10^{-3}$, and 226 1-qubit gates, each with an average error of $4.238\cdot10^{-4}$. This is for the estimation of the expectation value of a single observable, and therefore the procedure has to be performed three times, one for each observable of interest. Assuming that the error would grow as 
$e^{-n\varepsilon}$ where $n$ is the number of gates and $\varepsilon$ the average gate error, we expect an error of approximately $ 52 \% $ in the value of each observable. However, we see in figure \ref{fig5} that the use of error mitigation can significantly improve these poor values of the fidelity.

\section{Summary and conclusions}

 We use digital quantum computing to simulate the creation of particles by the dynamics of an expanding spacetime. By means of a boson-qubit mapping, we digitize a system consisting of two modes of a massive quantum scalar field in a spacetime undergoing homogeneous and isotropic expansion, transitioning from one stationary state to another through an inflationary period. We study the number of particles created after some interaction time as a function of the expansion rate, both by simulating the circuit and by actual experimental implementation on IBM quantum computers. The circuit consists of a few hundred of quantum gates. We find that state-of-the-art error mitigation techniques such as ZNE are useful to improve the estimation of the number of particles and the fidelity of the state.

Quantum computing is still in a very early stage. It is said that we are living in the NISQ era, in which quantum technology, such as computers, is not yet large enough in terms of qubit size to surpass the power of classical computers, and is too noisy to be used for reasonably long computations, like the one we have done in this work.

In terms of size, some of IBM's quantum processors, such as Condor with its 1,121 qubits, have already surpassed the threshold of what a classical computer can do. However, noise, as we have seen in this work, remains an unresolved issue.

Using only 4 qubits, we have built a circuit to simulate a quantum field placed on a gravitational background, limiting ourselves to just one excitation of the ground state. We could continue to include excited states by simply adding more qubits, but this would require many more gates, causing more noise and reducing fidelity, making it impossible to use the circuit for physical predictions.
One of the short-term goals in the field of quantum computing is precisely this: achieving a quantum simulation of a physical system that cannot be done with a classical computer. Unfortunately, this necessarily depends on improving the efficiency of existing quantum computers. In our four-qubit example the use of error mitigation techniques helped to alleviate the issue, as in a recent example with 127 qubits and only three layers of gates \cite{kim2023evidence}. Whether this would be still the case in a problem combining both large qubit number and large circuit depth remains an open issue. While we do not have unrestricted access to a state-of-the-art large quantum device at the moment, our preliminary few-qubit results can be seen as a first step towards this sort of experiments, as we discussed in more detail below.

Although the interaction that was presented here can be studied analytically, it would be computationally harder to do so with more modes or excitations per mode, which would essentially require the same techniques described here. For instance, with the currently available 127-qubit computers, we could consider more than 50 excitations per mode, which would also allow us to use a wider range of parameters. These experiments would presumably enter into the post-classical regime, since it is known that circuits with more than 100 qubits and enough circuit depth are computationally very hard for classical computers. However, even considering the improvements in reducing error rates, it is likely that it would not be enough to compensate for the dramatic improvement in the number of quantum gates that would be required to simulate such a large number of excitations. Therefore, as commented above, it would be necessary to check if ZNE could still render useful results for both large number of qubits and large circuit depth. Since we do not have unrestricted access to a large quantum computer at the moment, the question lies beyond the scope of this paper and remains open for future investigation. If answered successfully, it would show the utility of a quantum computer to realize a computation of fundamental interest beyond the typical approximations employed classically.  
% Other systems of quantum field theory in curved spacetime could also be studied. In this work, we have studied a simple example in which the vibration modes do not mix, so it is analytically solvable. However, these techniques can also be extrapolated to systems that are not analytically solvable, allowing us to study new physics that is currently unknown.

\begin{backmatter}

%\section*{Declarations}

\section*{Acknowledgements}%% if any
We thank Sieglinde Pfaendler for her help and support.
%\section*{Abbreviations}

%C. S.: Carlos Sabín

\section*{Funding}%% if any

We acknowledge the use of IBM Quantum Credits for this work. The views expressed are those of the authors, and do not reflect the official policy or position of IBM or the IBM Quantum team. 
C.S. acknowledges financial support through the Ramón y Cajal Programme (RYC2019-028014-I).

%\section*{Abbreviations}%% if any
%Text for this section\ldots
%
\section*{Data availability}%% if any
Data is provided within the manuscript or supplementary information files or available from the corresponding author upon reasonable request.
\section*{Ethics approval and consent to participate}%% if any
Not applicable.
\section*{Competing interests}
The authors declare that they have no competing interests.

\section*{Consent for publication}%% if any
Not applicable

\section*{Authors' contributions}
CS suggested and supervised the project. MDM made the computations, programmed and launched the experiments, analysed the data and wrote the first version of the manuscript. Both authors contributed to the final writing of the manuscript.

% \bibliographystyle{bmc-mathphys} % Style BST file (bmc-mathphys, vancouver, spbasic).
% \bibliography{bibliografia}      % Bibliography file (usually '*.bib' )

\newcommand{\BMCxmlcomment}[1]{}

\BMCxmlcomment{

<refgrp>

<bibl id="B1">
  <title><p>How far are we from the quantum theory of gravity?</p></title>
  <aug>
    <au><snm>Woodard</snm><fnm>R P</fnm></au>
  </aug>
  <source>Reports on Progress in Physics</source>
  <publisher>{IOP} Publishing</publisher>
  <pubdate>2009</pubdate>
  <volume>72</volume>
  <issue>12</issue>
  <fpage>126002</fpage>
  <url>https://doi.org/10.1088/0034-4885/72/12/126002</url>
</bibl>

<bibl id="B2">
  <title><p>Spin Entanglement Witness for Quantum Gravity</p></title>
  <aug>
    <au><snm>Bose</snm><fnm>S</fnm></au>
    <au><snm>Mazumdar</snm><fnm>A</fnm></au>
    <au><snm>Morley</snm><fnm>GW</fnm></au>
    <au><snm>Ulbricht</snm><fnm>H</fnm></au>
    <au><snm>Toro{\v{s} }</snm><fnm>M</fnm></au>
    <au><snm>Paternostro</snm><fnm>M</fnm></au>
    <au><snm>Geraci</snm><fnm>AA</fnm></au>
    <au><snm>Barker</snm><fnm>PF</fnm></au>
    <au><snm>Kim</snm><fnm>M. S.</fnm></au>
    <au><snm>Milburn</snm><fnm>G</fnm></au>
  </aug>
  <source>Physical Review Letters</source>
  <publisher>American Physical Society ({APS})</publisher>
  <pubdate>2017</pubdate>
  <volume>119</volume>
  <issue>24</issue>
  <url>https://doi.org/10.11032Fphysrevlett.119.240401</url>
</bibl>

<bibl id="B3">
  <title><p>Particle creation by black holes</p></title>
  <aug>
    <au><snm>Hawking</snm><fnm>S. W.</fnm></au>
  </aug>
  <source>Communications in Mathematical Physics</source>
  <pubdate>1975</pubdate>
  <volume>43</volume>
  <issue>3</issue>
  <fpage>199</fpage>
  <lpage>220</lpage>
  <url>https://doi.org/10.1007/BF02345020</url>
</bibl>

<bibl id="B4">
  <title><p>Notes on black-hole evaporation</p></title>
  <aug>
    <au><snm>Unruh</snm><fnm>W. G.</fnm></au>
  </aug>
  <source>Phys. Rev. D</source>
  <publisher>American Physical Society</publisher>
  <pubdate>1976</pubdate>
  <volume>14</volume>
  <fpage>870</fpage>
  <lpage>-892</lpage>
  <url>https://link.aps.org/doi/10.1103/PhysRevD.14.870</url>
</bibl>

<bibl id="B5">
  <title><p>Chronology protection conjecture</p></title>
  <aug>
    <au><snm>Hawking</snm><fnm>S. W.</fnm></au>
  </aug>
  <source>Phys. Rev. D</source>
  <publisher>American Physical Society</publisher>
  <pubdate>1992</pubdate>
  <volume>46</volume>
  <fpage>603</fpage>
  <lpage>-611</lpage>
  <url>https://link.aps.org/doi/10.1103/PhysRevD.46.603</url>
</bibl>

<bibl id="B6">
  <title><p>Quantized Fields and Particle Creation in Expanding Universes. I</p></title>
  <aug>
    <au><snm>Parker</snm><fnm>L</fnm></au>
  </aug>
  <source>Phys. Rev.</source>
  <publisher>American Physical Society</publisher>
  <pubdate>1969</pubdate>
  <volume>183</volume>
  <fpage>1057</fpage>
  <lpage>-1068</lpage>
  <url>https://link.aps.org/doi/10.1103/PhysRev.183.1057</url>
</bibl>

<bibl id="B7">
  <title><p>Quantized Fields and Particle Creation in Expanding Universes. II</p></title>
  <aug>
    <au><snm>Parker</snm><fnm>L</fnm></au>
  </aug>
  <source>Phys. Rev. D</source>
  <publisher>American Physical Society</publisher>
  <pubdate>1971</pubdate>
  <volume>3</volume>
  <fpage>346</fpage>
  <lpage>-356</lpage>
  <url>https://link.aps.org/doi/10.1103/PhysRevD.3.346</url>
</bibl>

<bibl id="B8">
  <title><p>Thermal radiation produced by the expansion of the Universe</p></title>
  <aug>
    <au><snm>Parker</snm><fnm>L</fnm></au>
  </aug>
  <source>Nature</source>
  <publisher>Nature Publishing Group UK London</publisher>
  <pubdate>1976</pubdate>
  <volume>261</volume>
  <issue>5555</issue>
  <fpage>20</fpage>
  <lpage>-23</lpage>
</bibl>

<bibl id="B9">
  <title><p>Observation of quantum Hawking radiation and its entanglement in an analogue black hole</p></title>
  <aug>
    <au><snm>Steinahuer</snm><fnm>J</fnm></au>
  </aug>
  <source>Nature Phys</source>
  <pubdate>2016</pubdate>
  <volume>12</volume>
  <fpage>959</fpage>
  <url>https://doi.org/10.1038/nphys3863</url>
</bibl>

<bibl id="B10">
  <title><p>Colloquium: Stimulating uncertainty: Amplifying the quantum vacuum with superconducting circuits</p></title>
  <aug>
    <au><snm>Nation</snm><fnm>P. D.</fnm></au>
    <au><snm>Johansson</snm><fnm>J. R.</fnm></au>
    <au><snm>Blencowe</snm><fnm>M. P.</fnm></au>
    <au><snm>Nori</snm><fnm>F</fnm></au>
  </aug>
  <source>Rev. Mod. Phys.</source>
  <publisher>American Physical Society</publisher>
  <pubdate>2012</pubdate>
  <volume>84</volume>
  <fpage>1</fpage>
  <lpage>-24</lpage>
  <url>https://link.aps.org/doi/10.1103/RevModPhys.84.1</url>
</bibl>

<bibl id="B11">
  <title><p>Quantum simulation of Unruh radiation</p></title>
  <aug>
    <au><snm>Hu</snm><fnm>J</fnm></au>
    <au><snm>Feng</snm><fnm>L</fnm></au>
    <au><snm>Zhang</snm><fnm>Z</fnm></au>
    <au><snm>Chin</snm><fnm>C</fnm></au>
  </aug>
  <source>Nature Physics</source>
  <publisher>Nature Publishing Group UK London</publisher>
  <pubdate>2019</pubdate>
  <volume>15</volume>
  <issue>8</issue>
  <fpage>785</fpage>
  <lpage>-789</lpage>
</bibl>

<bibl id="B12">
  <title><p>Analogue cosmological particle creation in an ultracold quantum fluid of light</p></title>
  <aug>
    <au><snm>Steinhauer</snm><fnm>J</fnm></au>
    <au><snm>Abuzarli</snm><fnm>M</fnm></au>
    <au><snm>Aladjidi</snm><fnm>T</fnm></au>
    <au><snm>Bienaim{\'e}</snm><fnm>T</fnm></au>
    <au><snm>Piekarski</snm><fnm>C</fnm></au>
    <au><snm>Liu</snm><fnm>W</fnm></au>
    <au><snm>Giacobino</snm><fnm>E</fnm></au>
    <au><snm>Bramati</snm><fnm>A</fnm></au>
    <au><snm>Glorieux</snm><fnm>Q</fnm></au>
  </aug>
  <source>Nature Communications</source>
  <publisher>Nature Publishing Group UK London</publisher>
  <pubdate>2022</pubdate>
  <volume>13</volume>
  <issue>1</issue>
  <fpage>2890</fpage>
</bibl>

<bibl id="B13">
  <title><p>Quantum field simulator for dynamics in curved spacetime</p></title>
  <aug>
    <au><snm>Viermann</snm><fnm>C</fnm></au>
    <au><snm>Sparn</snm><fnm>M</fnm></au>
    <au><snm>Liebster</snm><fnm>N</fnm></au>
    <au><snm>Hans</snm><fnm>M</fnm></au>
    <au><snm>Kath</snm><fnm>E</fnm></au>
    <au><snm>Parra L{\'o}pez</snm><fnm>{\'A}</fnm></au>
    <au><snm>Tolosa Sime{\'o}n</snm><fnm>M</fnm></au>
    <au><snm>S{\'a}nchez Kuntz</snm><fnm>N</fnm></au>
    <au><snm>Haas</snm><fnm>T</fnm></au>
    <au><snm>Strobel</snm><fnm>H</fnm></au>
    <au><cnm>others</cnm></au>
  </aug>
  <source>nature</source>
  <publisher>Nature Publishing Group UK London</publisher>
  <pubdate>2022</pubdate>
  <volume>611</volume>
  <issue>7935</issue>
  <fpage>260</fpage>
  <lpage>-264</lpage>
</bibl>

<bibl id="B14">
  <title><p>Quantum computing in the NISQ era and beyond</p></title>
  <aug>
    <au><snm>Preskill</snm><fnm>J</fnm></au>
  </aug>
  <source>Quantum</source>
  <publisher>Verein zur F{\"o}rderung des Open Access Publizierens in den Quantenwissenschaften</publisher>
  <pubdate>2018</pubdate>
  <volume>2</volume>
  <fpage>79</fpage>
</bibl>

<bibl id="B15">
  <title><p>Quantum supremacy using a programmable superconducting processor</p></title>
  <aug>
    <au><snm>Arute</snm><fnm>F</fnm></au>
    <au><snm>Arya</snm><fnm>K</fnm></au>
    <au><snm>Babbush</snm><fnm>R</fnm></au>
    <au><snm>Bacon</snm><fnm>D</fnm></au>
    <au><snm>Bardin</snm><fnm>JC</fnm></au>
    <au><snm>Barends</snm><fnm>R</fnm></au>
    <au><snm>Biswas</snm><fnm>R</fnm></au>
    <au><snm>Boixo</snm><fnm>S</fnm></au>
    <au><snm>Brandao</snm><fnm>FG</fnm></au>
    <au><snm>Buell</snm><fnm>DA</fnm></au>
    <au><cnm>others</cnm></au>
  </aug>
  <source>Nature</source>
  <publisher>Nature Publishing Group</publisher>
  <pubdate>2019</pubdate>
  <volume>574</volume>
  <issue>7779</issue>
  <fpage>505</fpage>
  <lpage>-510</lpage>
</bibl>

<bibl id="B16">
  <title><p>Strong quantum computational advantage using a superconducting quantum processor</p></title>
  <aug>
    <au><snm>Wu</snm><fnm>Y</fnm></au>
    <au><snm>Bao</snm><fnm>WS</fnm></au>
    <au><snm>Cao</snm><fnm>S</fnm></au>
    <au><snm>Chen</snm><fnm>F</fnm></au>
    <au><snm>Chen</snm><fnm>MC</fnm></au>
    <au><snm>Chen</snm><fnm>X</fnm></au>
    <au><snm>Chung</snm><fnm>TH</fnm></au>
    <au><snm>Deng</snm><fnm>H</fnm></au>
    <au><snm>Du</snm><fnm>Y</fnm></au>
    <au><snm>Fan</snm><fnm>D</fnm></au>
    <au><cnm>others</cnm></au>
  </aug>
  <source>Physical review letters</source>
  <publisher>APS</publisher>
  <pubdate>2021</pubdate>
  <volume>127</volume>
  <issue>18</issue>
  <fpage>180501</fpage>
</bibl>

<bibl id="B17">
  <title><p>Beyond quantum supremacy: the hunt for useful quantum computers</p></title>
  <aug>
    <au><snm>Brooks</snm><fnm>M</fnm></au>
  </aug>
  <source>Nature</source>
  <publisher>Nature Publishing Group</publisher>
  <pubdate>2019</pubdate>
  <volume>574</volume>
  <issue>7776</issue>
  <fpage>19</fpage>
  <lpage>-22</lpage>
</bibl>

<bibl id="B18">
  <title><p>Efficient Variational Quantum Simulator Incorporating Active Error Minimization</p></title>
  <aug>
    <au><snm>Li</snm><fnm>Y</fnm></au>
    <au><snm>Benjamin</snm><fnm>SC</fnm></au>
  </aug>
  <source>Physical Review X</source>
  <pubdate>2017</pubdate>
  <volume>7</volume>
</bibl>

<bibl id="B19">
  <title><p>Error Mitigation for Short-Depth Quantum Circuits</p></title>
  <aug>
    <au><snm>Temme</snm><fnm>K</fnm></au>
    <au><snm>Bravyi</snm><fnm>S</fnm></au>
    <au><snm>Gambetta</snm><fnm>JM</fnm></au>
  </aug>
  <source>Physical Review Letters</source>
  <publisher>American Physical Society (APS)</publisher>
  <pubdate>2017</pubdate>
  <volume>119</volume>
  <issue>18</issue>
</bibl>

<bibl id="B20">
  <title><p>Error mitigation extends the computational reach of a noisy quantum processor</p></title>
  <aug>
    <au><snm>Kandala</snm><fnm>A</fnm></au>
    <au><snm>Temme</snm><fnm>K</fnm></au>
    <au><snm>Córcoles</snm><fnm>AD</fnm></au>
    <au><snm>Mezzacapo</snm><fnm>A</fnm></au>
    <au><snm>Chow</snm><fnm>JM</fnm></au>
    <au><snm>Gambetta</snm><fnm>JM</fnm></au>
  </aug>
  <source>Nature</source>
  <publisher>Springer Science and Business Media LLC</publisher>
  <pubdate>2019</pubdate>
  <volume>567</volume>
  <issue>7749</issue>
  <fpage>491–495</fpage>
</bibl>

<bibl id="B21">
  <title><p>Evidence for the utility of quantum computing before fault tolerance</p></title>
  <aug>
    <au><snm>Kim</snm><fnm>Y</fnm></au>
    <au><snm>Eddins</snm><fnm>A</fnm></au>
    <au><snm>Anand</snm><fnm>S</fnm></au>
    <au><snm>Wei</snm><fnm>KX</fnm></au>
    <au><snm>Van Den Berg</snm><fnm>E</fnm></au>
    <au><snm>Rosenblatt</snm><fnm>S</fnm></au>
    <au><snm>Nayfeh</snm><fnm>H</fnm></au>
    <au><snm>Wu</snm><fnm>Y</fnm></au>
    <au><snm>Zaletel</snm><fnm>M</fnm></au>
    <au><snm>Temme</snm><fnm>K</fnm></au>
    <au><cnm>others</cnm></au>
  </aug>
  <source>Nature</source>
  <publisher>Nature Publishing Group UK London</publisher>
  <pubdate>2023</pubdate>
  <volume>618</volume>
  <issue>7965</issue>
  <fpage>500</fpage>
  <lpage>-505</lpage>
</bibl>

<bibl id="B22">
  <title><p>Quantum fields in curved space</p></title>
  <aug>
    <au><snm>Birrel</snm><fnm>N. D.</fnm></au>
    <au><snm>Davies</snm><fnm>P. C. W.</fnm></au>
  </aug>
  <publisher>Cambridge University Press</publisher>
  <section><title><p>3.4 Cosmological particle creation: a simple example</p></title></section>
  <pubdate>1982</pubdate>
  <fpage>59</fpage>
  <lpage>62</lpage>
</bibl>

<bibl id="B23">
  <title><p>Über die Krümmung des Raumes</p></title>
  <aug>
    <au><snm>Friedmann</snm><fnm>A</fnm></au>
  </aug>
  <source>Zeitschrift für Physik</source>
  <pubdate>1922</pubdate>
  <volume>10</volume>
  <fpage>377</fpage>
  <lpage>386</lpage>
</bibl>

<bibl id="B24">
  <title><p>Über die Möglichkeit einer Welt mit konstanter negativer Krümmung des Raumes</p></title>
  <aug>
    <au><snm>Friedmann</snm><fnm>A</fnm></au>
  </aug>
  <source>Zeitschrift für Physik</source>
  <pubdate>1924</pubdate>
  <volume>21 (1)</volume>
  <fpage>326</fpage>
  <lpage>332</lpage>
</bibl>

<bibl id="B25">
  <title><p>Un Univers homogène de masse constante et de rayon croissant rendant compte de la vitesse radiale des nébuleuses extra-galactiques</p></title>
  <aug>
    <au><snm>Lemaître</snm><fnm>G</fnm></au>
  </aug>
  <source>Annales de la Société Scientifique de Bruxelles</source>
  <pubdate>1927</pubdate>
  <volume>47</volume>
  <fpage>49</fpage>
  <lpage>59</lpage>
</bibl>

<bibl id="B26">
  <title><p>Kinematics and World-Structure</p></title>
  <aug>
    <au><snm>Robertson</snm><fnm>HP</fnm></au>
  </aug>
  <source>Astrophysical Journal</source>
  <pubdate>1935</pubdate>
  <volume>82</volume>
  <fpage>284</fpage>
</bibl>

<bibl id="B27">
  <title><p>Kinematics and World-Structure II</p></title>
  <aug>
    <au><snm>Robertson</snm><fnm>HP</fnm></au>
  </aug>
  <source>Astrophysical Journal</source>
  <pubdate>1936</pubdate>
  <volume>83</volume>
  <fpage>187</fpage>
</bibl>

<bibl id="B28">
  <title><p>Kinematics and World-Structure III</p></title>
  <aug>
    <au><snm>Robertson</snm><fnm>HP</fnm></au>
  </aug>
  <source>Astrophysical Journal</source>
  <pubdate>1936</pubdate>
  <volume>83</volume>
  <fpage>257</fpage>
</bibl>

<bibl id="B29">
  <title><p>On Milne's Theory of World-Structure</p></title>
  <aug>
    <au><snm>Walker</snm><fnm>AG</fnm></au>
  </aug>
  <source>Proceedings of the London Mathematical Society</source>
  <pubdate>1937</pubdate>
  <volume>s2-42</volume>
  <fpage>90</fpage>
  <lpage>127</lpage>
</bibl>

<bibl id="B30">
  <title><p>Quantentheorie und fünfdimensionale Relativitätstheorie</p></title>
  <aug>
    <au><snm>Klein</snm><fnm>O</fnm></au>
  </aug>
  <source>Zeitschrift für Physik</source>
  <pubdate>1926</pubdate>
  <volume>37</volume>
  <fpage>895–906</fpage>
</bibl>

<bibl id="B31">
  <title><p>Der Comptoneffekt nach der Schr\"odingerschen Theorie</p></title>
  <aug>
    <au><snm>Gordon</snm><fnm>W.</fnm></au>
  </aug>
  <source>Zeitschrift für Physik</source>
  <pubdate>1926</pubdate>
  <volume>40</volume>
  <fpage>117</fpage>
  <lpage>133</lpage>
</bibl>

<bibl id="B32">
  <title><p>On a new method in the theory of superconductivity</p></title>
  <aug>
    <au><snm>Bogoljubov</snm><fnm>N. N.</fnm></au>
    <au><snm>Tolmachov</snm><fnm>V. V.</fnm></au>
    <au><snm>Širkov</snm><fnm>D. V.</fnm></au>
  </aug>
  <source>Fortschritte der Physik</source>
  <pubdate>1958</pubdate>
  <volume>6</volume>
  <fpage>605</fpage>
  <lpage>682</lpage>
</bibl>

<bibl id="B33">
  <title><p>Comments on the theory of superconductivity</p></title>
  <aug>
    <au><snm>Valatin</snm><fnm>J. G.</fnm></au>
  </aug>
  <source>Nuovo Cimento</source>
  <pubdate>1958</pubdate>
  <volume>7</volume>
  <fpage>843–857</fpage>
</bibl>

<bibl id="B34">
  <title><p>Black hole explosions?</p></title>
  <aug>
    <au><snm>Hawking</snm><fnm>SW</fnm></au>
  </aug>
  <source>Nature</source>
  <pubdate>1974</pubdate>
  <volume>248</volume>
  <fpage>30</fpage>
  <lpage>31</lpage>
</bibl>

<bibl id="B35">
  <title><p>Notes on black-hole evaporation</p></title>
  <aug>
    <au><snm>Unruh</snm><fnm>WG</fnm></au>
  </aug>
  <source>Physical Review D</source>
  <pubdate>1976</pubdate>
  <volume>14</volume>
  <fpage>870</fpage>
  <lpage>892</lpage>
</bibl>

<bibl id="B36">
  <title><p>New formalism for two-photon quantum optics. II. Mathematical foundation and compact notation</p></title>
  <aug>
    <au><snm>Schumaker</snm><fnm>BL</fnm></au>
    <au><snm>Caves</snm><fnm>CM</fnm></au>
  </aug>
  <source>Phys. Rev. A</source>
  <publisher>American Physical Society</publisher>
  <pubdate>1985</pubdate>
  <volume>31</volume>
  <fpage>3093</fpage>
  <lpage>-3111</lpage>
  <url>https://link.aps.org/doi/10.1103/PhysRevA.31.3093</url>
</bibl>

<bibl id="B37">
  <title><p>Obtainment of thermal noise from a pure quantum state</p></title>
  <aug>
    <au><snm>Yurke</snm><fnm>B.</fnm></au>
    <au><snm>Potasek</snm><fnm>M.</fnm></au>
  </aug>
  <source>Phys. Rev. A</source>
  <publisher>American Physical Society</publisher>
  <pubdate>1987</pubdate>
  <volume>36</volume>
  <fpage>3464</fpage>
  <lpage>-3466</lpage>
  <url>https://link.aps.org/doi/10.1103/PhysRevA.36.3464</url>
</bibl>

<bibl id="B38">
  <title><p>Quantum Simulations of Physics Problems</p></title>
  <aug>
    <au><snm>Somma</snm><fnm>R</fnm></au>
    <au><snm>Ortiz</snm><fnm>G</fnm></au>
    <au><snm>Knill</snm><fnm>E</fnm></au>
    <au><snm>Gubernatis</snm><fnm>J</fnm></au>
  </aug>
  <source>International Journal of Quantum Information</source>
  <pubdate>2003</pubdate>
  <volume>1 (2)</volume>
  <fpage>189</fpage>
  <lpage>206</lpage>
</bibl>

<bibl id="B39">
  <title><p>Digital Quantum Simulation of Linear and Nonlinear Optical Elements</p></title>
  <aug>
    <au><snm>Sabín</snm><fnm>C</fnm></au>
  </aug>
  <source>Quantum Reports</source>
  <pubdate>2020</pubdate>
  <volume>2</volume>
  <fpage>208–220</fpage>
</bibl>

<bibl id="B40">
  <title><p>Digital quantum simulation of beam splitters and squeezing with IBM quantum computers</p></title>
  <aug>
    <au><snm>Encinar</snm><fnm>PC</fnm></au>
    <au><snm>Agustí</snm><fnm>A</fnm></au>
    <au><snm>Sabín</snm><fnm>C</fnm></au>
  </aug>
  <source>Physical Review A</source>
  <pubdate>2021</pubdate>
  <volume>104</volume>
</bibl>

<bibl id="B41">
  <title><p>Digital quantum simulation of quantum gravitational entanglement with IBM quantum computers</p></title>
  <aug>
    <au><snm>Sabín</snm><fnm>C</fnm></au>
  </aug>
  <source>EPJ Quantum Technology</source>
  <pubdate>2023</pubdate>
  <volume>10</volume>
</bibl>

<bibl id="B42">
  <title><p>Digital quantum simulation of gravitational optomechanics with IBM quantum computers</p></title>
  <aug>
    <au><snm>Rufo</snm><fnm>PGC</fnm></au>
    <au><snm>Mazumdar</snm><fnm>A</fnm></au>
    <au><snm>Bose</snm><fnm>S</fnm></au>
    <au><snm>Sabín</snm><fnm>C</fnm></au>
  </aug>
  <source>EPJ Quantum Technology</source>
  <pubdate>2024</pubdate>
  <volume>11</volume>
</bibl>

</refgrp>
} % end of \BMCxmlcomment

%\printbibliography[heading=bibintoc]
% \section*{Figures}

\end{backmatter}
\end{document}